\documentclass[useAMS,usenatbib]{mn2e}

\usepackage{graphicx}

\title[Electron capture of strongly screening nuclides $^{56}$Fe, $^{56}$Co, $^{56}$Ni , $^{56}$Mn ,$^{56}$Cr and $^{56}$V in presupernova]{Electron capture of strongly screening nuclides $^{56}$Fe, $^{56}$Co, $^{56}$Ni , $^{56}$Mn ,$^{56}$Cr and $^{56}$V in presupernova}
\author[Liu Jing-Jing ]{Liu Jing-Jing$^{1}$\thanks{E-mail:liujingjing68@126.com}
\footnotemark[1]\thanks{Project supported by the Advanced Academy Special Foundation of Sanya under Grant No 2011YD14.}\\
$^{1}$College of Science and Tecnology, Qiongzhou University, Sanya, 572022, China}
\begin{document}

\date{Accepted 2012 September 15. Received 2012 December 14; in original form 2012 August 11}

\pagerange{\pageref{firstpage}--\pageref{lastpage}} \pubyear{2012}

\maketitle

\label{firstpage}

\begin{abstract}
According to the Shell-Model Monte Carlo method, basing
on the Random Phase Approximation and the linear response theory, we
carried out an estimation on electron capture of strongly screening
nuclides $^{56}$Fe, $^{56}$Co, $^{56}$Ni , $^{56}$Mn ,$^{56}$Cr and
$^{56}$V in strong electron screening (SES)in presupernova. The
 EC rates are decreased greatly  and even exceed 21.5\% in SES.  We also compare our results with those of Aufderheide
(AFUD), which calculated by the method of Aufderheide in SES.
Our results are agreed reasonably well with AUFD at higher
density-temperature surroundings (e.g. $\rho_7>60, T_9=15.40$) and the maximum error is $\sim
0.5$\%. However, the maximum error is $\sim 13.0$\% at lower density surroundings (e.g. $^{56}$Cr at $\rho_7=10,
T_9=15.40, Y_e=0.41$ ). On the other hand, we also compared  our results in SES with those of FFN's and Nabi's, which
is in the case without SES. The comparisons show that our results are lower more than one order magnitude than FFN's, but about
7.23\% than Nabi's.
\end{abstract}

\begin{keywords}
 stars: supernovae, Physical Date and Processes: nuclear reactions.
\end{keywords}

\section{Introduction}

The electron capture (EC) rates of $^{56}$Fe, $^{56}$Co, $^{56}$Ni,
$^{56}$Mn, $^{56}$Cr and $^{56}$V play key roles in the final
evolution of massive stars, especially for presupernova evolution.
Some pioneer works on EC rates are investigated by
\citet{b6}(FFN); \citet{b1}(AUFD); \citet{b11, b12}; and \citet{b24} under supernova explosion
conditions. \citet{b13, b14, b15, b16, b17, b18} also discussed the
weak interaction reactions on these nuclides. But their discussions did not
consider the influence of strong electron screening (SES) on EC
rates.

The SES has been raised a strong interest among nuclear astrophysicist. The SES has also always been an interest issue and challenging problem for stellar weak-interaction rates in presupernova stellar evolution and nucleosynthesis. It is extremely interesting, important and necessary for us to understand, solve and calculate accurately the SES and screening corrections in dense stars under the condition of the relativistic degenerate electron liquid.

In the process of EC, what role will play on earth by SES in stars?
How does SES affect on the EC rates? These problems have already
been discussed by \citet{b7}; \citet{b2}; \citet{b19, b20}; \citet{b13, b21,
b22, b23}; and \citet{b9}. \citet{b10} have improved previous rate evaluations by properly including screening corrections to the reaction rates into account. Their researches show that it is
extremely important and necessary to calculate accurately the screening corrections to the EC rates in dense stars.

Due to the importance of SES in astrophysical surroundings,
according to the Shell-Model Monte Carlo (SMMC) method, which
discussed amply by \citet{b5}, in this paper, basing on
the linear response theory \cite{b9},  we investigate the affection on EC rates of these iron nuclides
by SES. We also discuss the electron capture cross section with Random Phase Approximation theory (RPA) and the rate of
change of electron fraction (RCEF) due to EC in SES. On the other
hand, we also discuss the error factor $C_{1}$, which is
comparisions of the rates of $\lambda^{s}_{SMMC}$ which calculated
by the method of SMMC with those of $\lambda^{s}_{AUFD}$, which
calculate by the method of \citet{b1} in SES and the screening factors $C_{2}$
in and not in SES. We also present the comparisons of our results in SES with those of FFN's
and Nabi's, which is in the case without SES. The present paper is organized as follows: in the next section, we analyses the EC rates
in SES in stellar interiors. Some numerical results and discussion
are given in Section 3. And some conclusions are summarized in Section 4.

\section[]{The EC in SES in stellar interiors}

The stellar electron capture rates for the $k$ th nucleus (Z, A) in
thermal equilibrium at temperature $T$ is given by a sum over the
initial parent states $i$ and the final daughter states $f$ \citep{b6,
b1}
\begin{equation}
\label{eq.1}
  \lambda_{k}=\lambda_{ec}=\sum_{i}\frac{2(J_i+1)e^{\frac{-E_i}{kT}}}{G(Z,A,T)} \sum_{f}\lambda_{if}
\end{equation}
The EC
rate from one of the initial states to all possible final states is
$\lambda_{if}$. The $J_i$ and $E_i$ are the spin and excitation energies of the
parent states,  $G(Z,A,T)$ is the nuclear partition function and given by
\begin{equation}
\label{eq.2}
G(Z,A,T)=\sum_i(2J_i+1)exp(-\frac{E_i}{kT})
\end{equation}

using the level density formula, $\vartheta(E, J, \pi)$ , the contribution from the excite states is discussed. Thus the nuclear partition function approximately becomes \citep{b1}
\begin{eqnarray}
\label{eq.3}
G(Z,A,T) &\approx& (2J_0+1)+ \int_0^\infty dE \int _{J,\pi}dJd\pi (2J_i+1)\nonumber\\
&&\times \vartheta(E, J, \pi)exp(-\frac{E_i}{kT})
\end{eqnarray}
where the level density  is given by \citep{b8, b26}
\begin{equation}
\label{eq.4}
\vartheta(E, J, \pi)=\frac{1}{\sqrt{2\pi}\sigma}\frac{\sqrt{\pi}}{12a^{\frac{1}{4}}}\times\frac{exp[2\sqrt{a(E-\delta)}]}{(E-\delta)^{\frac{5}{4}}} f(E, J, \pi)
\end{equation}
where
\begin{equation}
\label{eq.5}
f(E, J, \pi)=\frac{1}{2}\frac{(2J+1)}{2\sigma^2}exp[-\frac{J(J+1)}{2\sigma^2}]
\end{equation}
where $a$ is the level density parameter, $\delta$ is the backshift (pairing correction).  $\sigma$ is defined as
\begin{equation}
\label{eq.6}
\sigma=(\frac{2m_uAR^2}{2\hbar^2})^{\frac{1}{2}}[\frac{(E-\delta)}{a}]^{\frac{1}{4}}
\end{equation}
where $R$ is the radius and $m_u=\frac{1}{N_A}$ is the atomic mass unit.

Based on the RPA theory with a global parameterization of the
single particle numbers, the EC rates is related to the electron
capture cross-section  by \citep{b10}
\begin{equation}
\label{eq.2}
\lambda_{if}=\frac{1}{\pi^2\hbar^3}\sum_{if}\int^{\infty}_{\varepsilon_0}p^2_e\sigma_{ec}(\varepsilon_n,\varepsilon_i,\varepsilon_f)f(\varepsilon_n,U_F,T)d\varepsilon_n
\end{equation}
where $\varepsilon_0=\max(Q_{if}, 1)$. $p_e=\sqrt{\varepsilon_n-1}$
is the momentum of the incoming electron with energy
$\varepsilon_{n}$ and $U_{F}$ is the electron chemical potential,
$T$ is the electron temperature. Note that in this paper all of the
energies and the moments are respectively in units of $m_e c^2$ and
$m_e c$, where $m_e$ is the electron mass and $c$ is the light
speed in vacuum. The phase space factor is defined as
\begin{equation}
  f=f(\varepsilon_{n},U_F,T)=[1+exp(\frac{\varepsilon_{n}-U_F}{kT})]^{-1}
\label{eq.3}
\end{equation}
where a electron with the energy $\varepsilon_n$ from an initial
proton single particle state with energy $\varepsilon_i$ to a
neutron single particle state with energy $\varepsilon_f$. Due to
the energy conservation, the electron, proton and neutron energies
are related to the neutrino energy, and $Q$-value for the capture
reaction\citep{b3}
\begin{equation}
  Q_{i,f}=\varepsilon_{e}-\varepsilon_{\nu}=\varepsilon_{n}-\varepsilon_{\nu}=\varepsilon^{n}_{f}-\varepsilon^{p}_{i}
\label{eq.4}
\end{equation}
and we have
\begin{equation}
  \varepsilon^{n}_{f}-\varepsilon^{p}_{i}=\varepsilon^{\ast}_{if}+\hat{\mu}+\Delta_{np}
\label{eq.5}
\end{equation}
where $\hat{\mu}=\mu_{n}-\mu_p$, the difference between neutron and
proton chemical potentials in the nucleus and
$\Delta_{np}=M_{n}c^2-M_{p}c^2=1.293Mev$, the neutron and the proton
mass difference. $Q_{00}=M_{f}c^2-M_{i}c^2=\hat{\mu}+\Delta_{np}$,
with $M_{i}$ and $M_{f}$ being the masses of the parent nucleus and
the daughter nucleus respectively; $\varepsilon^{\ast}_{if}$
corresponds to the excitation energies in the daughter nucleus at
the states of the zero temperature.

The electron chemical potential is found by inverting the expression
for the lepton number density
\begin{equation}
  n_e=\frac{\rho}{\mu_e} =\frac{8\pi}{(2\pi)^3}\int^\infty_0 p^2_e(f_{-e}-f_{+e})dp_e
\label{eq.6}
\end{equation}
where $\rho$ is the density in $g/cm^3$, $\mu_e$ is the average
molecular weight. $\lambda_e=\frac{h}{m_{e}c}$ is the Compton
wavelength,
$f_{-e}=[1+exp(\frac{\varepsilon_{n}-U_{F}-1}{kT})]^{-1}$ and
$f_{-e}=[1+exp(\frac{\varepsilon_{n}-U_{F}+1}{kT})]^{-1}$ are the
electron and positron distribution functions respectively, $k$ is
the Boltzmann constant.

According to the Shell-Model Monte Carlo method, which discussed the
Gamow-Teller strength distributions, the total cross section by EC
is given by \citep{b10}
\begin{eqnarray}
\sigma_{ec}&=& \sigma_{ec}(\varepsilon_n)=\sum_{if}\frac{(2J_{i}+1)exp(-\beta E_i)}{Z_A}\sigma_{fi}(Ee)\nonumber\\
&=&\sum_{if}\frac{(2J_{i}+1)exp(-\beta E_i)}{Z_A}\sigma_{fi}(En) \nonumber\\
&=& 6g^{2}_{wk}\int d\xi(\varepsilon_{n}-\xi)^2 \frac{G^2_A}{12\pi}
S_{GT^+}(\xi) F(Z,\varepsilon_n)
\label{eq.7}
\end{eqnarray}
where $g_{wk}=1.1661\times 10^{-5}\rm{Gev^{-2}}$ is the weak
coupling constant and $G_A$ is the axial vector form-factor which at
zero momentum is $G_A=1.25$.  The $\varepsilon_n$ is the total rest
mass and kinetic energies; $F(Z, \varepsilon_n)$ is the Coulomb wave
correction which is the ratio of the square of the electron wave
function distorted by the coulomb scattering potential to the square
of wave function of the free electron.

$S_{GT^+}$ is the total amount of Gamow-teller(GT) strength
available for an initial state is given by summing over a complete
set o final states in Gamow-teller transition matrix elements
$|M_{GT}|^{2}_{if}$.The SMMC method is also used to
calculate the response function $R_A(\tau)$ of an operator $\hat{A}$
at an imaginary-time $\tau$. By using a spectral distribution of
initial and final states $|i\rangle$ and $|f\rangle$ with energies
$E_i$ and $E_f$. $R_A(\tau)$ is given by \citep{b11}
\begin{equation}
R_A(\tau)=\frac{\sum_{if}(2J_i+1)e^{-\beta E_i}e^{-\tau
(E_f-E_i)}|\langle f|\hat{A}|i\rangle|^2}{\sum_i (2J_i+1)e^{-\beta
E_i}}
\label{eq:008}
\end{equation}
Note that the total strength for the operator is given by
$R(\tau=0)$. The strength distribution is given by
\begin{eqnarray}
S_{GT^+}(E) &=& \frac{\sum_{if}\delta (E-E_f+E_i)(2J_i+1)e^{-\beta
E_i}|\langle f|\hat{A}|i\rangle|^2}{\sum_i (2J_i+1)e^{-\beta E_i}} \nonumber\\
&=& S_{A}(E)
 \label{eq:009}
\end{eqnarray}
which is related to $R_A(\tau)$ by a Laplace Transform,
$R_A(\tau)=\int_{-\infty}^{\infty}S_A(E)e^{-\tau E}dE$. Note that
here $E$ is the energy transfer within the parent nucleus, and that
the strength distribution $S_{GT^+}(E)$ has units of $\rm
{Mev^{-1}}$ and $\beta=\frac{1}{T_N}$, $T_N$ is the nuclear
temperature.

The presupernova EC rates is given by folding the total cross
section with the flux of a degenerate relativistic electron gas in
the case without SES \citep{b11}
\begin{eqnarray}
\lambda^{0}_{ec}&=&\frac{\ln2}{6163}\int^{\infty}_{0}d\xi S_{GT}\frac{c^3}{(m_{e}c^2)^5}\nonumber\\
&& \int^{\infty}_{p_0}dp_{e}p^2_e(-\xi+\varepsilon_n)
F(Z,\varepsilon_n)f
\label{eq.8}
\end{eqnarray}
The $p_0$ is defined as
\begin{equation}
p_0=\left\lbrace \begin{array}{ll}~\sqrt{Q^2_{if}-1}~~~~~~( Q_{if}<-1)\\
                                  ~0 ~~~~~~(otherwise).
                             \end{array} \right.
\label{eq.9}
\end{equation}

Using the linear response theory, \citet{b9} calculated
the screening potential for relativistic degenerate electrons. A
more precise screening potential is given by
\begin{equation}
D=7.525\times10^{-3}z(\frac{10z\rho_7}{A})^{\frac{1}{3}}J(r_s,R)
(\rm{Mev}) \label{eq.11}
\end{equation}
where $\rho_7$ is the density in units of $10^7\rm{g/cm^3}$,
$J(r_s,R)$, $r_s$ and $R$ can be found in Ref. \citep{b9}. The formula
(12) is valid for $10^{-5}\leq r_s \leq 10^{-1}, 0\leq R\leq 50$
conditions, which are usually fulfilled in the pre-supernova
environment.

If the electron is strongly screened and the screening energy is
high enough in order not to be neglected in high density plasma. Its
energy will decrease from $\varepsilon$ to
$\varepsilon^{'}=\varepsilon-D$ in the decay reaction due to
electron screening. At the same time, the screening relatively
decreases the number of high energy electrons with energies higher
than the threshold energy for electron capture. The threshold energy
increases from $\varepsilon_0$ to $\varepsilon_s=\varepsilon_0+D$.
Thus the EC rates with SES becomes
\begin{eqnarray}
\label{eq.10}
\lambda^{s}_{ec}&=&\frac{\ln2}{6163}\int^{\infty}_{0}d\xi S_{GT^+}\frac{c^3}{(m_{e}c^2)^5}\nonumber\\[1mm]
&&
\int^{\infty}_{p_0}dp_{e}p^2_e(-\xi+\varepsilon_n) F(Z,\varepsilon_n)f(\varepsilon_n,U_F,T)\nonumber\\[1mm]
&=&\frac{\ln2}{6163}\int^{\infty}_{0}d\xi S_{GT^+}\frac{c^3}{(m_{e}c^2)^5}\nonumber\\[1mm]
&&
\int^{\infty}_{\varepsilon_s}d\varepsilon^{'}\varepsilon{'}(\varepsilon^{'2}-1)^{\frac{1}{2}}(-\xi+\varepsilon^{'})^2
F(Z,\varepsilon^{'})f
\end{eqnarray}

We define the error factors $C_{1}$, which compare our results of
$\lambda^s_{SMMC}$, which discussed by method of SMMC with those
of $\lambda^s_{AFUD}$, which calculated by the method of AUFD. We also define the screening factors $C_{2}$ with and without SES.
\begin{equation}
C_{1}=\frac{(\lambda^s_{SMMC}-\lambda^s_{AFUD})}{\lambda^s_{SMMC}}
\label{eq.12}
\end{equation}

\begin{equation}
C_{2}=\frac{\lambda^s_{SMMC}}{\lambda^0_{SMMC}}
\label{eq.13}
\end{equation}

On the other hand, the RCEF plays a key role in stellar
evolution and presupernova outburst. In order to understand how
would the SES effect on RCEF, the RCEF due to EC reaction on the $k$
th nucleus in SES is defined as
\begin{equation}
\dot{Y^{ec}_{e}}(k)=-\frac{X_k}{A_k}\lambda^{s}_k
\label{eq.14}
\end{equation}
where $\lambda^{s}_k$ is the EC rates in SES; $X_k$ is the mass
fraction of the $k$ th nucleus and $A_k$ is the mass number of the
$k$ th nucleus.

\section{Some numerical results and discussion}

Figure 1 shows some numerical results on EC rates at
$T_9=3.40,Y_e=0.47$ and $T_9=15.40,Y_e=0.41$ in SES. ($T_9$ is the
temperature in units of $10^9$ K). We find the SES has different
effects on EC at different density and temperatures. The EC rates
are increased greatly and even exceed by seven orders of
magnitude at lower temperature (e.g. $T_9=3.40,Y_e=0.47$ for
$^{56}$Cr). The lower the temperature and the higher the density, the larger
the influence on EC is. Because the electron energy is so low at
lower temperatures and the SES potential is so high in higher
density that the SES can strongly affect the EC rates. On the other
hand, with increasing of the density, there are different affections
on EC for different nuclide. It is caused by different Q-values and
the transition orbits. For example,the Q-values of nuclides $^{56}$Fe,
$^{56}$Mn, $^{56}$Cr and $^{56}$V are negative, but the others are
plus (e. g. $Q_0=4.06Mev$ ; $Q_0=1.62Mev$ for $^{56}$Co and
$^{56}$Ni respectively). The RCEF is very sensitivity parameter in
EC process. From Figs 2, we find the SES effect largely on RCEF. The RCEF reduces greatly and even exceed for 7 orders of
magnitude in SES.

\begin{figure}
\centering
    \includegraphics[width=4cm,height=4cm]{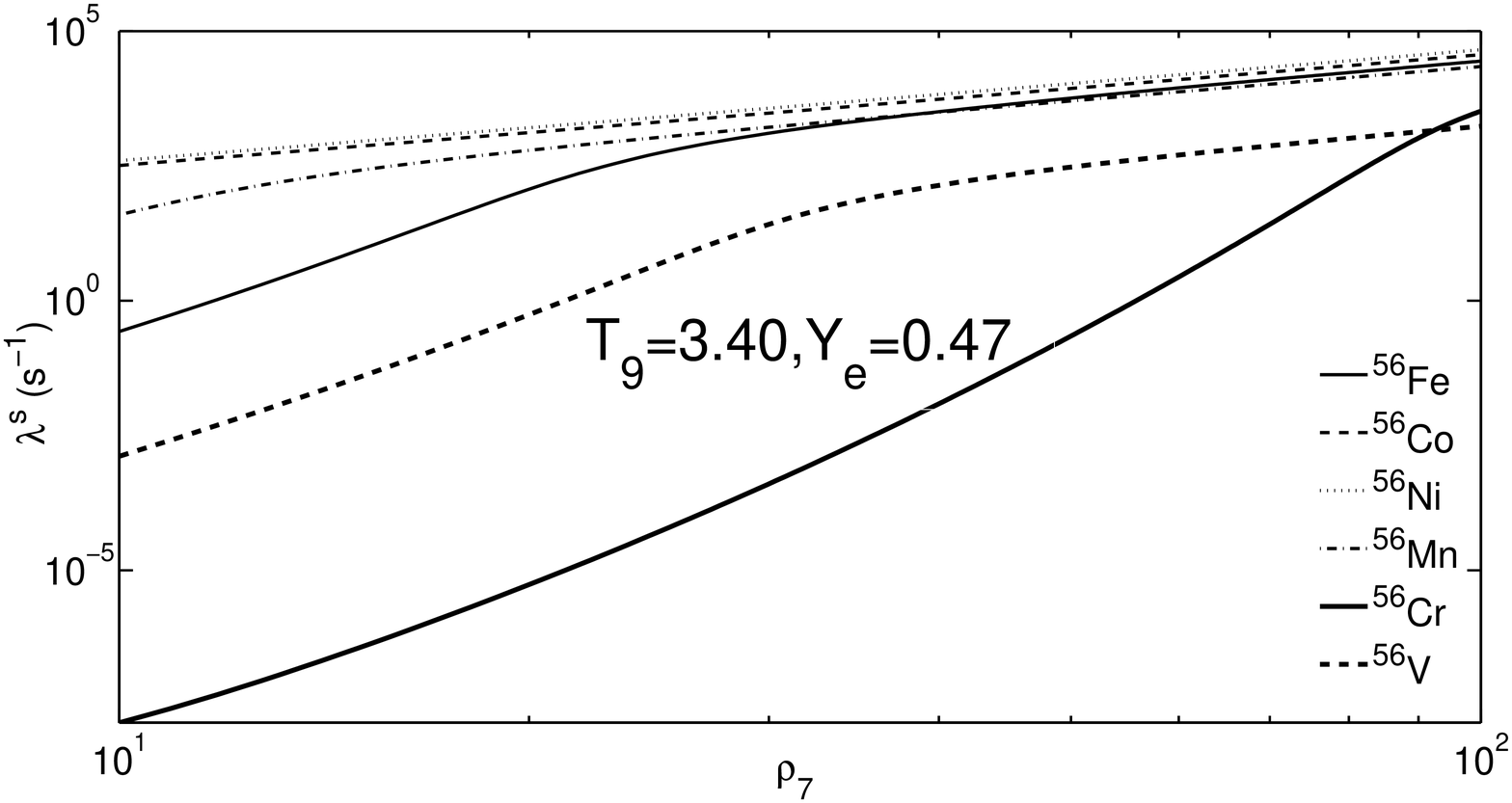}
    \includegraphics[width=4cm,height=4cm]{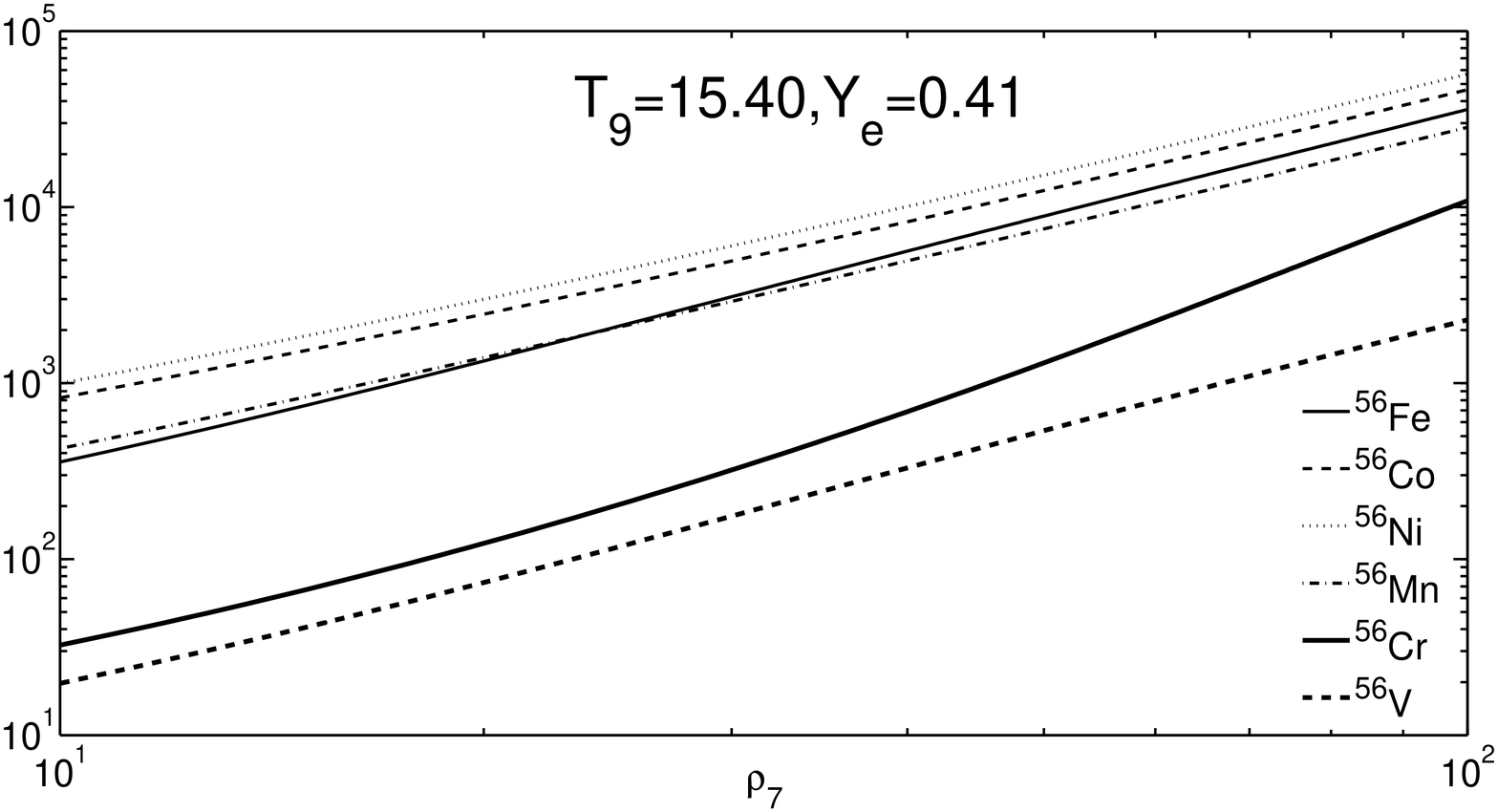}
    \caption{The EC rates as a function of $\rho_7$ at $T_9=3.40,Y_e=0.47$ and $T_9=15.40,Y_e=0.41$ in SES }
   \label{Fig:1}
\end{figure}

%
\begin{figure}
\centering
    \includegraphics[width=4cm,height=4cm]{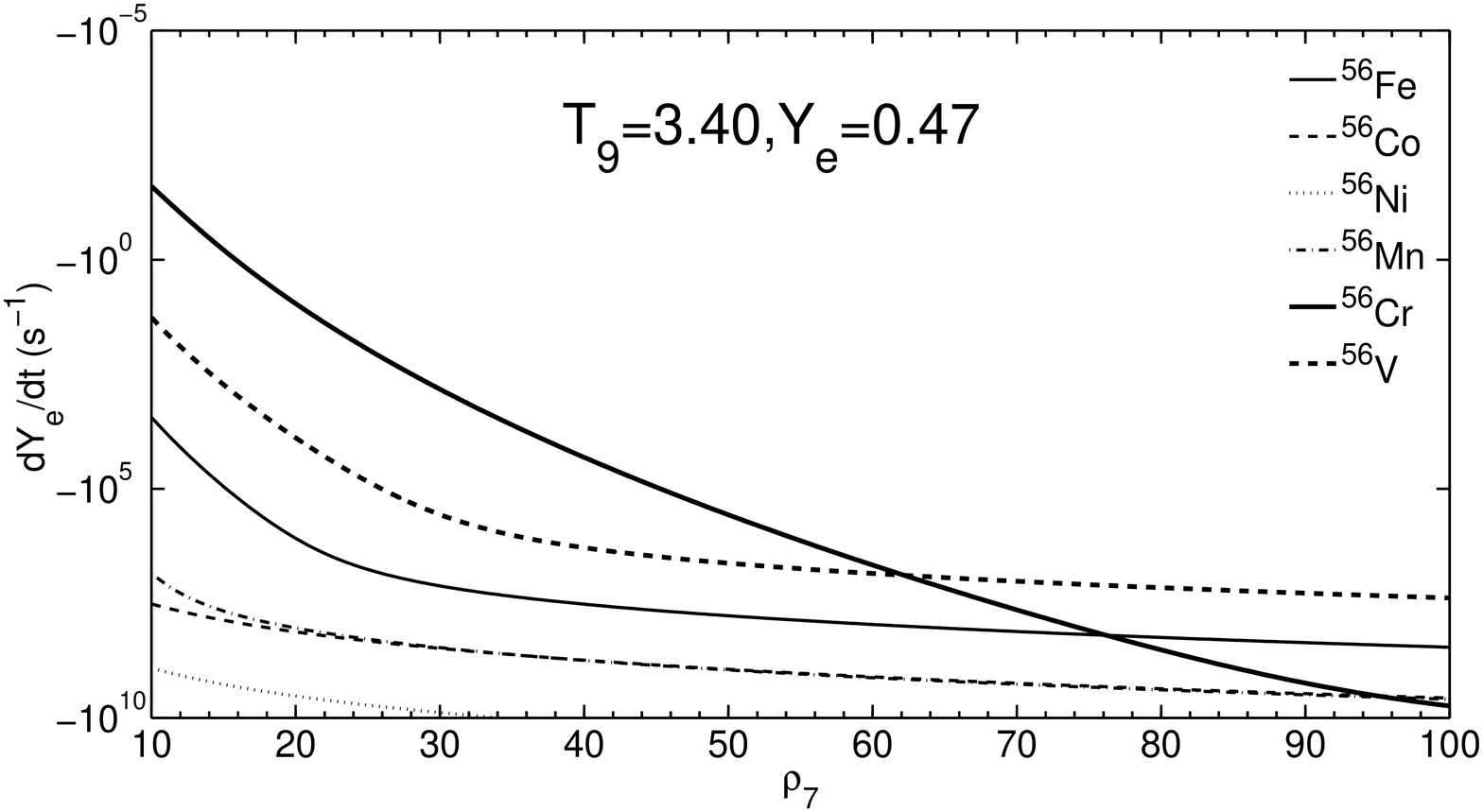}
    \includegraphics[width=4cm,height=4cm]{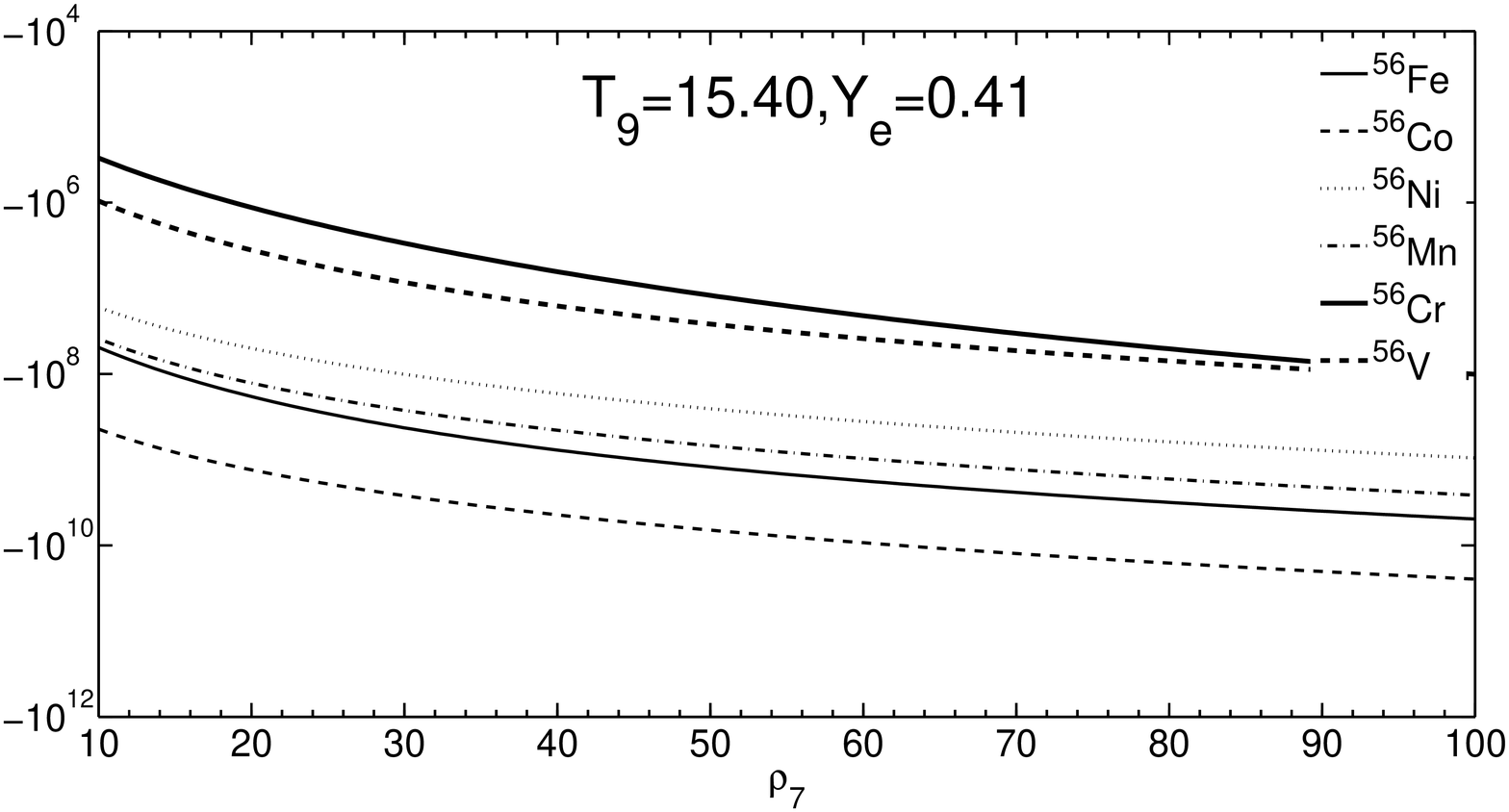}
   \caption{The $\dot{Y^{ec}_{e}}(k)$(RCEF) due to EC process as a function of $\rho_7$ at $T_9=3.40,Y_e=0.47$ and $T_9=15.40,Y_e=0.41$ in SES}
   \label{Fig:2}
\end{figure}
%
\begin{figure}
\centering
    \includegraphics[width=4cm,height=4cm]{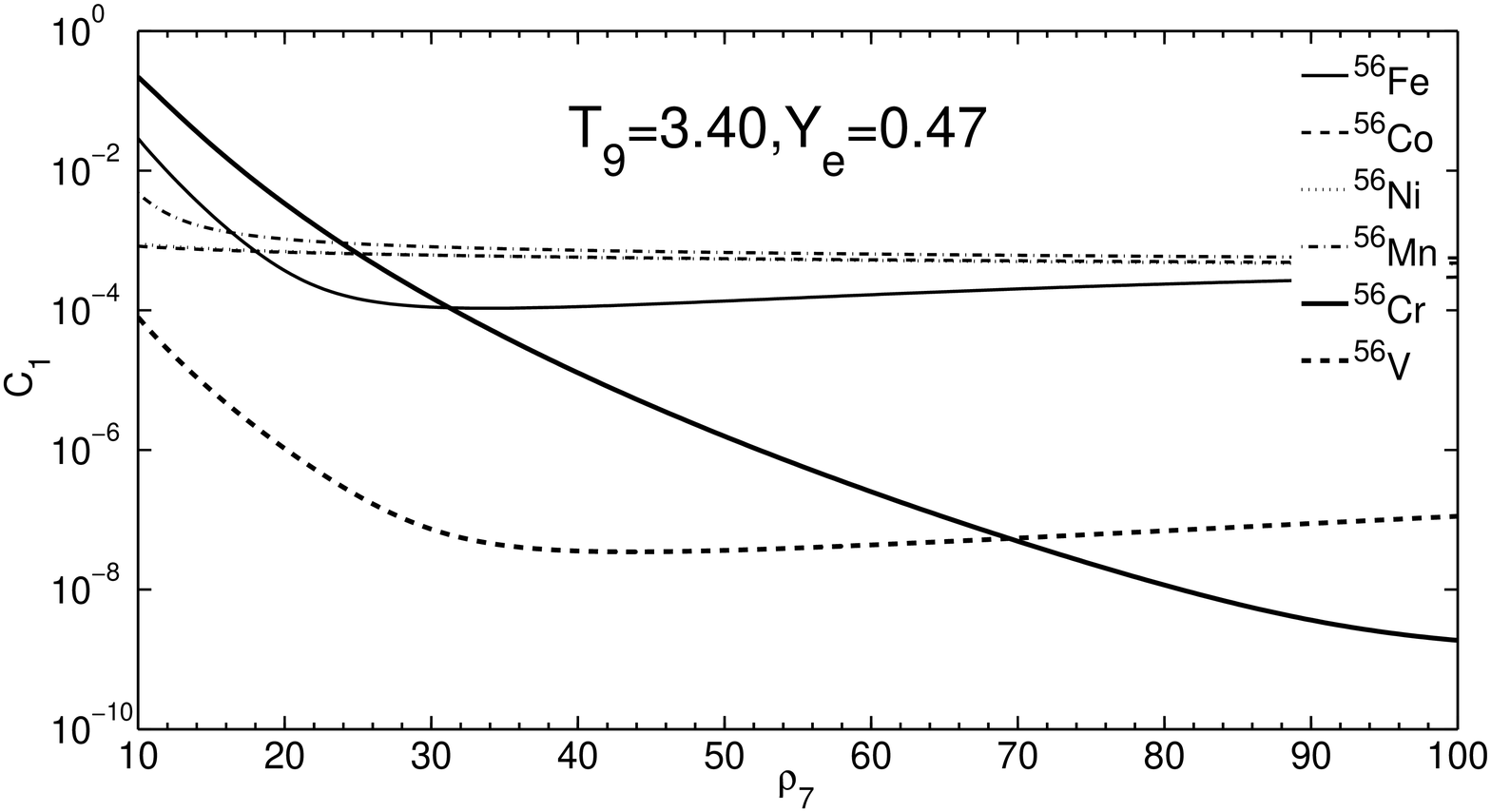}
    \includegraphics[width=4cm,height=4cm]{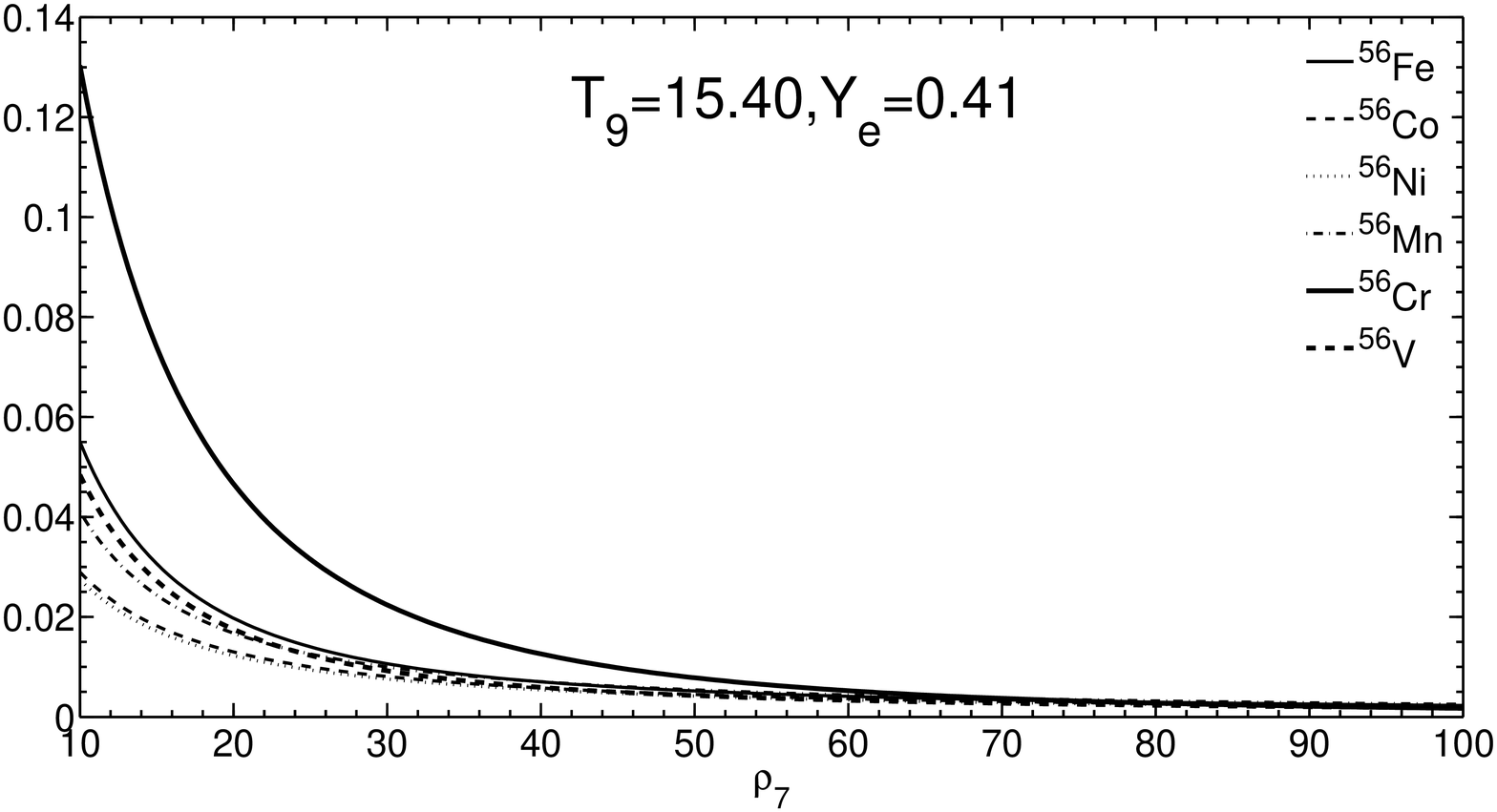}
   \caption{The error factor $C_{1}$ as a function of $\rho_7$ at $T_9=3.40,Y_e=0.47$ and $T_9=15.40,Y_e=0.41$ in SES}
   \label{Fig:3}
\end{figure}

\begin{figure}
\centering
    \includegraphics[width=4cm,height=4cm]{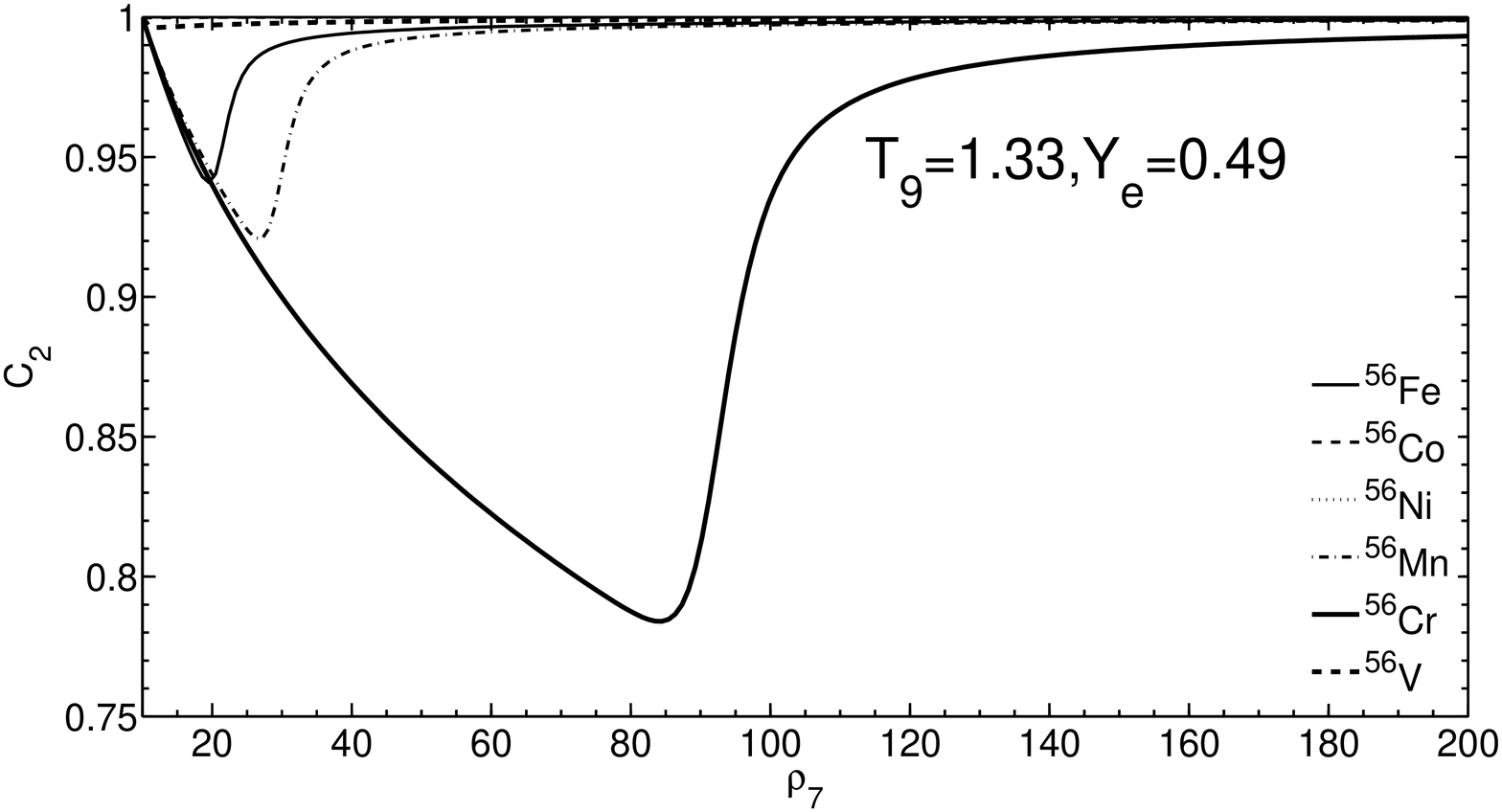}
    \includegraphics[width=4cm,height=4cm]{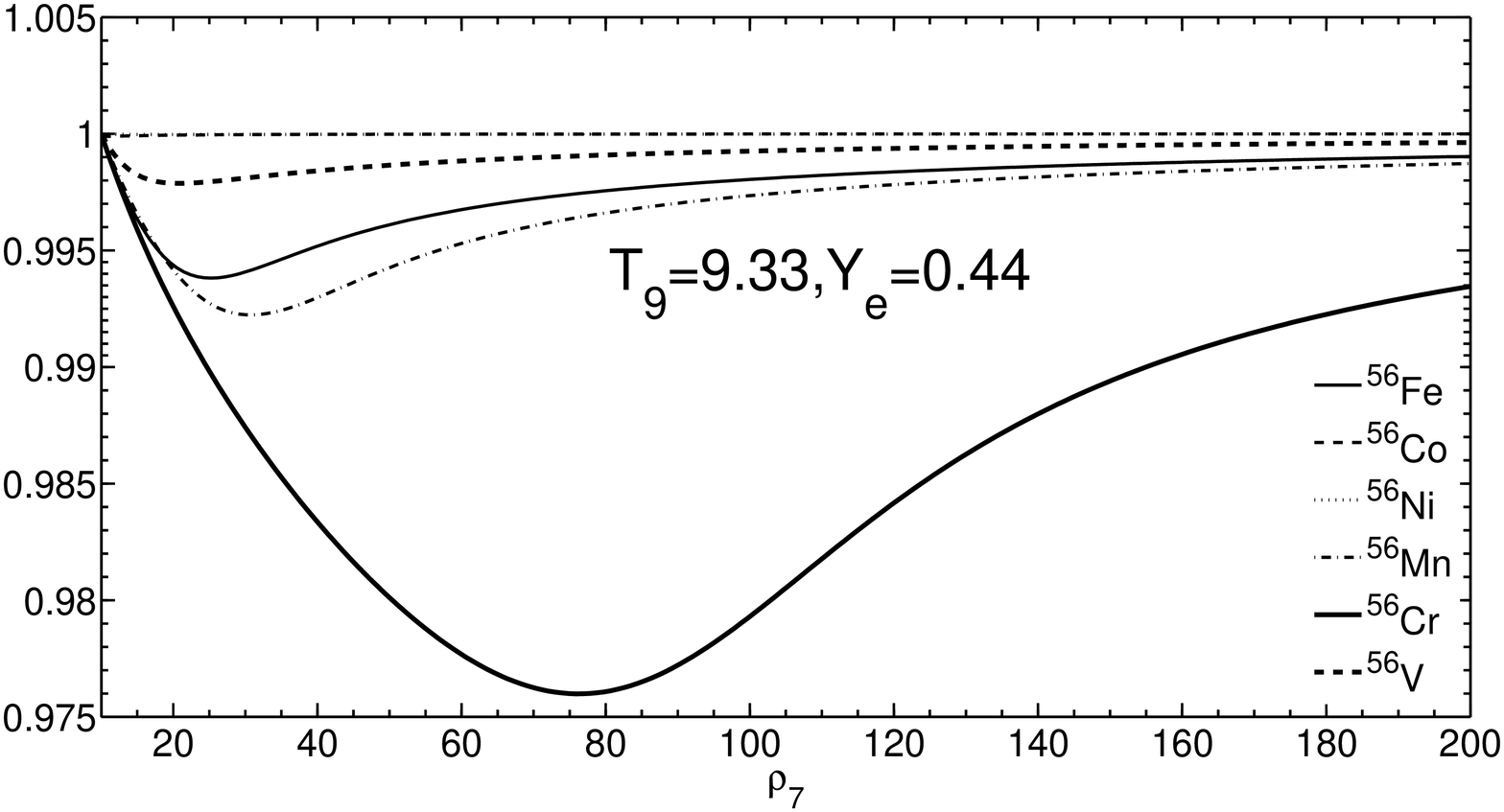}
   \caption{The screening factor $C_{2}$ as a function of $\rho_7$ at $T_9=1.33,Y_e=0.49$ and $T_9=9.33,Y_e=0.44$ in SES}
   \label{Fig:4}
\end{figure}

\begin{figure}
\centering
    \includegraphics[width=4cm,height=4cm]{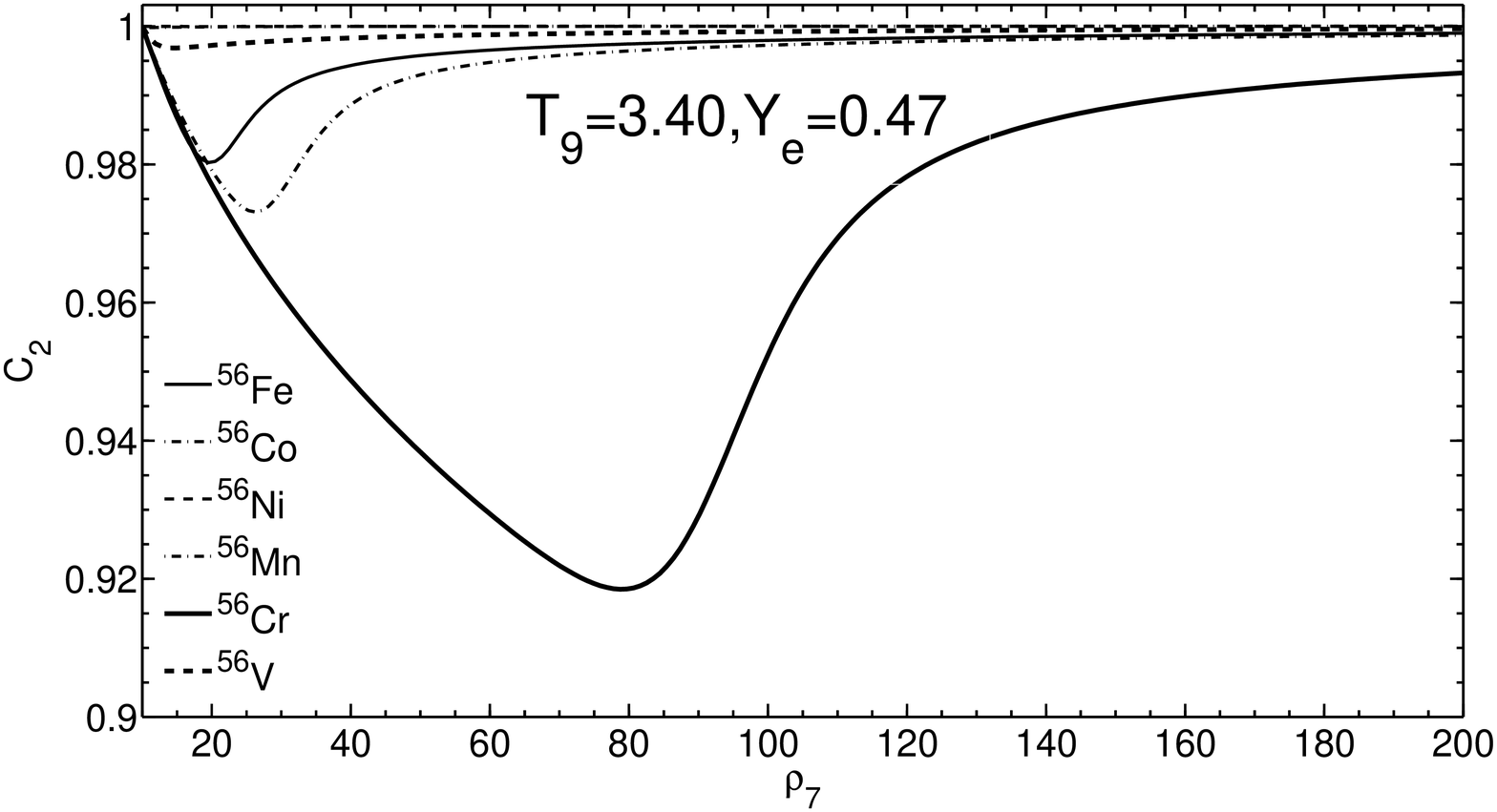}
    \includegraphics[width=4cm,height=4cm]{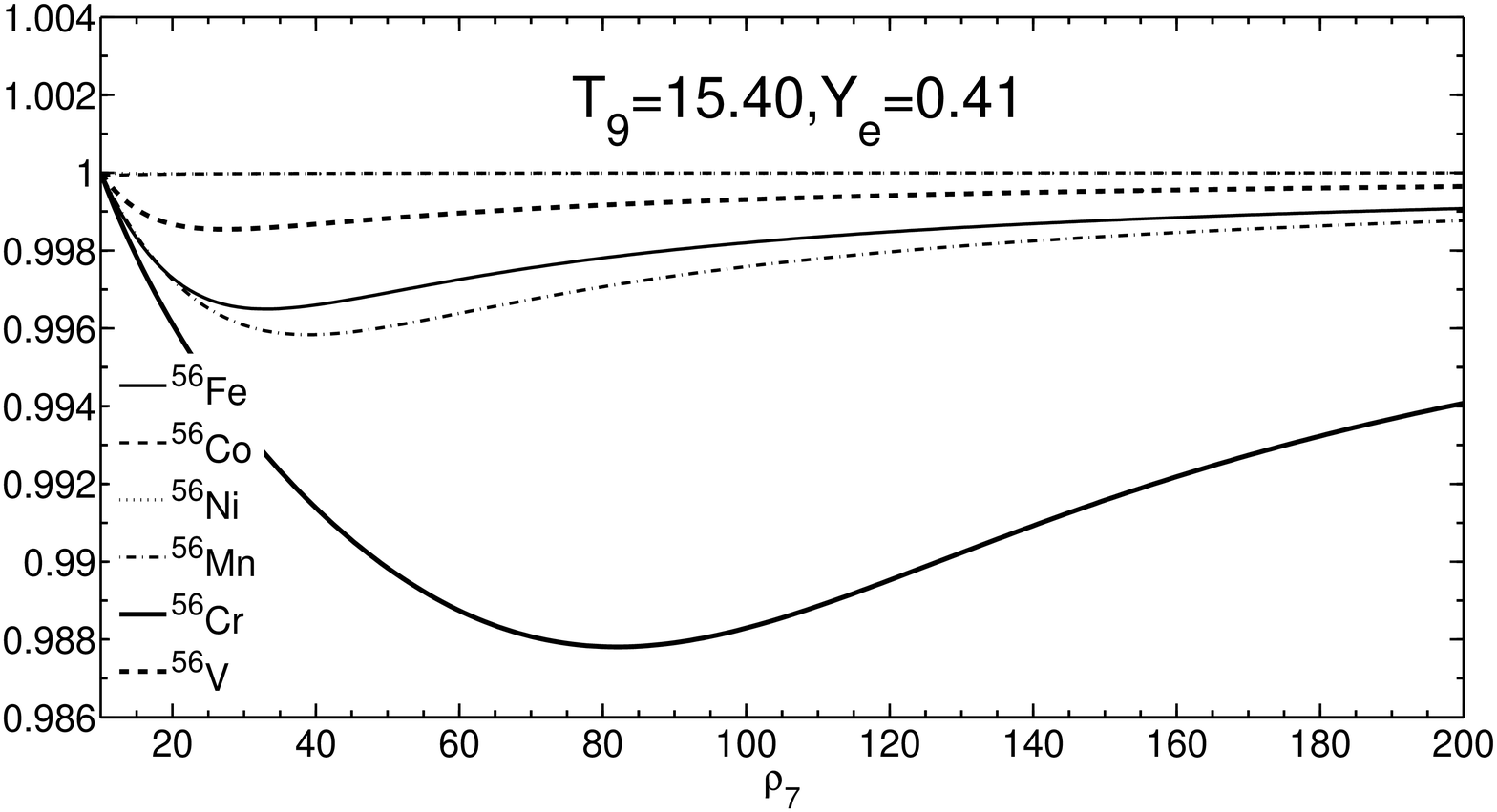}
   \caption{The screening factor $C_{2}$ as a function of $\rho_7$ at $T_9=3.40,Y_e=0.47$ and $T_9=15.40,Y_e=0.41$ in SES}
   \label{Fig:5}
\end{figure}

The error factors $C_{1}$ is plotted as a function of $\rho_7$ in
figure 3. We find the error factor decreases with increasing of $\rho_7$. The higher the temperature, the larger the affection on
factors $C_{1}$ is. According to our numerical calculations, one can
see that at lower temperature (e.g. $T_9=3.40,Y_e=0.47$ ) and higher
density-temperature($\rho_7>60; T_9=15.40,Y_e=0.41$), the fit is fairly good for two results and the
maximum error is $\sim 0.5$\%. However, the error is $\sim 5.50$ \%; $\sim 2.90$ \%; $\sim 2.70$
\%; $\sim 4.10$ \%; $\sim 13.0$ \%; $\sim 4.80$ \% for $^{56}$Fe,
$^{56}$Co, $^{56}$Ni , $^{56}$Mn, $^{56}$Cr and $^{56}$V at $\rho_7=10.0; T_9=15.40,Y_e=0.41$ respectively.

Basing on nuclear shell model, AUFD analyzed the nuclear excited
level by a Simple calculation on the nuclear excitation level
transitions in their works. The capture rates are made up of the
lower energy transition rates between the ground states and the
higher energy transition rates between GT resonance states. Some
research shows the work of AUFD is an oversimplification and
therefore the accuracy is limited. The charge exchange reactions (p, n) and (n, p) make it possible to observe, in principle, the total GT strength distribution in nuclei. The experimental information is particularly rich for some iron nuclides and the availability of both $\rm{GT}^+$ and $\rm{GT}^-$ makes it possible to study in detail the problem of renormalization of $\sigma\tau$ operators. We have calculated the total GT strength in a full p-f shell calculation, resulting in $\rm{B}(\rm{GT})=g_A^2|\langle\vec{\sigma}\tau_{+}\rangle|^2$, where $g_{\rm{A}}^2$ is axial-vector coupling constant. For example, under the conditions of presupernova the electron capture on $^{56}$Ni is dominated by the wave functions of the parent and daughter states and effected greatly in SES due to the fact that the electron screening potential can change the Coulomb wave function of electrons. And the total GT strength for $^{56}$Ni in a full p-f shell calculation, resulting in $\rm{B}(\rm{GT})=10.1 g_A^2$. The total GT strength of the other important nuclide $^{56}$Fe in a full p-f shell calculation can be found in the Ref. \citep{b4}. An average of the GT strength distribution is in fact obtained by SMMC method. A reliable replication of the GT distribution in the nucleus is carried out and detailed analysis by using an amplification of the electronic shell model. Thus the method is relative accuracy.

The screening factors $C_{2}$ is plotted as a function of $\rho_7$
in figure 4 and 5. We find the effect on EC rates is very obvious by SES.
The EC rates are reduced greatly and even exceed $\sim 21.5$\% and $\sim 8$\% in Fig.4 and 5 respectively. One can see that the lower the temperature, the larger the effect on EC rates is in SES. The SES mainly decreased the number of higher energy electrons joining the EC reaction. On the other hand, one can also see from Fig.4 and 5 that the screening factor is nearly the same at higher density and independent of
the temperature and density. The reason is that at higher density surroundings the electron energy is mainly determined by its Fermi energy,which is
strongly decided by density.
On the other hand, the lower the temperature, the larger the effect
on $C_{2}$ is. This is because the higher temperature, the higher the average
electron energy is, but the lower the SES potential is. In addition, because of
the smaller electron screening potential at the low density the lower the density, the smaller the effect is. As the density
increases, the $C_{2}$ increases gradually due to the increases of the shielding potential in EC reaction. As the density
further increases, the factor $C_{2}$ decreases and will be closed to identical at relative higher density. This is because the electron energy is mainly determined by Fermi energy at higher density and the effect is relatively weaker by temperature. As the density increases, the
electronic Fermi and shielding potential increases. The ratio
between shielding potential and Fermi energy has nothing to do with
density approximatively.


\begin{table}

\caption{The comparisons of our calculations in SES for nuclides
$^{56}$Fe, $^{56}$Co, $^{56}$Ni , $^{56}$Mn, $^{56}$Cr and $^{56}$V
with those of FFN's and Nabi's, which is in the case without SES at
$\rho Y_e=10^7 g/cm^3, T_9=3$. The ratio computes as
$k_1=\frac{\lambda^s_{LJ}}{\lambda^0_{ec}(\rm{FFN})}$ and
$k_2=\frac{\lambda^s_{LJ}}{\lambda^0_{ec}(\rm{Nabi})}$.}
\label{t.lbl}
\begin{center}
\tiny
\begin{tabular}{lccccr}
\hline\noalign{\smallskip}
Nuclide & $\lambda^0_{ec}$(FFN)  & $\lambda^0_{ec}$(Nabi) & $\lambda^s_{LJ}$  & $k_1$   & $k_2$   \\
 \hline\noalign{\smallskip}
$^{56}$Fe  & 5.236e-8  & 1.028e-9   & 1.013e-9     & 1.9347e-2  & 0.98541 \\ 
$^{56}$Co  & 0.0115    & 0.0032     & 0.00307      & 0.26696    & 0.95937  \\
$^{56}$Ni  & 0.0019    & 0.0013     & 0.00124      & 0.65262    & 0.95380 \\
$^{56}$Mn  & 4.140e-7  & 3.0903e-6  & 3.0368e-6    & 7.33530    & 0.98270  \\
$^{56}$Cr  & 2.460e-19 & 1.002e-16  & 9.872e-17    & 401.301    & 0.98521 \\ 
$^{56}$V   & 1.247e-14 & 7.178e-13  & 6.827e-13    & 54.7474    & 0.95110  \\
\noalign{\smallskip}\hline
\end{tabular}
\end{center}
\end{table}


\begin{table}
\tiny
\caption{The comparisons of our calculations in SES for nuclides
$^{56}$Fe, $^{56}$Co, $^{56}$Ni , $^{56}$Mn, $^{56}$Cr and $^{56}$V
with those of FFN's and Nabi's, which is in the case without SES at
$\rho Y_e=10^{11} g/cm^3, T_9=3$. The ratio computes as
$k_1=\frac{\lambda^s_{LJ}}{\lambda^0_{ec}(\rm{FFN})}$ and
$k_2=\frac{\lambda^s_{LJ}}{\lambda^0_{ec}(\rm{Nabi})}$.}
\label{t.lbl}
\begin{center}
\tiny
\begin{tabular}{lccccr}
\hline\noalign{\smallskip}
Nuclide & $\lambda ^0_{ec}$(FFN)  & $\lambda^0_{ec}$(Nabi) & $\lambda^s_{LJ}$  & $k_1$   & $k_2$   \\
 \hline\noalign{\smallskip}
$^{56}$Fe  & 5.408e4  & 1.683e4  & 1.5916e4    & 0.29430    & 0.94570 \\ 
$^{56}$Co  & 1.596e5  & 4.730e4  & 4.4510e4    & 0.27890    & 0.94100  \\
$^{56}$Ni  & 1.718e5  & 6.210e4  & 5.7610e4    & 0.33533    & 0.92770\\
$^{56}$Mn  & 1.574e4  & 1.089e5  & 1.0367e5    & 6.58640    & 0.95200  \\
$^{56}$Cr  & 1.189e4  & 5.960e3  & 5.6220e3    & 0.47280    & 0.94330 \\ 
$^{56}$V   & 1.862e4  & 4.860e3  & 4.6251e3    & 0.24840    & 0.95170  \\
\noalign{\smallskip}\hline
\end{tabular}
\end{center}
\end{table}

From the oxygen shell burning phase up to the end of convective core
silicon burning phase of massive stars the EC rates on these
nuclides play important roles. FFN had done some pioneer works
on EC rates. In order to understand how much the affection on EC is
by SES, the comparisons of our results ($\lambda^s_{LJ}$ ) in SES
with those of FFN's ( $\lambda^0_{ec}$(FFN))\citep{b6} and Nabi's
($\lambda^0_{ec}$ (Nabi))\citep{b24} in the case of without SES are
presented in a tabular form. Table 1 and 2 show the comparison of
our results in SES with those FFN's and Nabi's at $\rho Y_e=10^7
g/cm^3, T_9=3$ and $\rho Y_e=10^{11} g/cm^3, T_9=3$ respectively.

The calculated rates for most nuclides in SES are decreased and even
exceeded as much as by one orders of magnitude of compared to FFN's
results in the case without SES. The two tables also show the
comparisons of our results in SES with those of Nabi's, which
based on pn-QRPA theory without SES. The calculated rates for most
nuclides due to SES are decreased and even exceeded 4\% at $\rho
Y_e=10^7 g/cm^3, T_9=3$ (e. g. $^{56}$Co, $^{56}$Ni and
$^{56}$V).But the decrease is about 7.23\% at $\rho Y_e=10^{11}
g/cm^3, T_9=3$ for $^{56}$Ni.

According to the method of SMMC, Basing on RPA and linear response
theory, we have discussed the EC rates in SES. From above
calculations, we find the affection on EC by SES is obvious. The
comparisons show that the difference is larger between ours and
FFN's. On the other hand, our results in SES are generally lower
than Nabi's. The cause may shows as follows:  the electron
screening potential can change the Coulomb wave function of
electrons. The electron screening potential also decreases the
energy of the electron joining the EC reaction and generally
decreases the EC rates due to the screening increase the energy of
atomic nucleus in reaction. Moreover, SES evidently
decreases the number of higher- energy electrons of which the energy
is more than the threshold in EC reaction. So, the SES relatively
increases the threshold of reaction and also obviously decreases the
EC rates.

On the other hand, the electron capture of these neutrons rich
nuclides do not has measured mass, so that the EC Q-value has to be
estimated with a mass formal by FFN. FFN used the \cite{b25} Semiempirical atomic mass formula, Thus, the Q-value
used in the effective rates are quite different. Moreover, FFN did
not take into effect the process of particle emission from excited
states. FFN adopts the so-called Brink's hypothesis in their
calculations. This hypothesis assumes that the GT strength
distribution on excited states is the same as for the ground state,
only shifted by the excitation energy of the state. Their work
simplifies the nuclear excited energy level transition calculation
method. Therefore, the calculation method is a little rough and the
larger difference appears in the comparisons.

Using the pn-QRPA theory, Nabi expanded the FFN's works and analyzed
nuclear excitation energy distribution. They had taken into
consideration the particle emission processes, which constrain the
parent excitation energies. However in the GT transitions
considered in their works, only low angular momentum states are
considered. The method of SMMC is actually draws an average of GT
intensity distribution of electron capture, the calculated results
are in good agreement with experiments, but the results slants
generally small, especially for some odd-A nuclides.

In summary, by analyzing the effect on EC rates due to SES, one can see that the SES has an evident effect
on EC rates for different nuclides, particularly for heavier nuclides whose threshold is negative at higher density. According to above calculations and discussion, one can conclude that the EC rates are
decreased greatly and even exceed ~21.5\%.

\section{Conclusions}

According to the method of SMMC, Basing on RPA and linear response
theory, we have discussed the EC rates of $^{56}$Fe, $^{56}$Co,
$^{56}$Ni , $^{56}$Mn, $^{56}$Cr and $^{56}$V in SES in
presupernova. We find the EC rates decreased greatly by SES and even
exceed $\sim 21.5$\% (e. g. $T_9=1.33,Y_e=0.49$ for $^{56}$Cr). The
lower the temperature, the larger the effect on EC is. The
$\dot{Y^{ec}_{e}}(k)$ is very sensitivity to SES and reduces
greatly, even exceed 7 orders of magnitude. We also compare
our results with those of AFUD. The error factor $C_{1}$ shows that
ours is agreed reasonably well with AUFD under the lower temperature
(e.g. $T_9=3.40,Y_e=0.47$ ) and higher density-temperature
surroundings (e.g. the maximum error is $\sim 0.5 $\% for
$\rho_7>60; T_9=15.40,Y_e=0.41$). However, the error is $\sim 5.50$
\%; $\sim 2.90$ \%; $\sim 2.70$ \%; $\sim 4.10$ \%; $\sim 13.0$ \%;
$\sim 4.80$ \% for $^{56}$Fe, $^{56}$Co, $^{56}$Ni , $^{56}$Mn,
$^{56}$Cr and $^{56}$V at $\rho_7=10.0; T_9=15.40,Y_e=0.41$
respectively. On the other hand, we compared our results in SES with
those of FFN's and Nabi's, which is in the case without SES. The
comparisons show that the difference is larger between ours and
FFN's. But, our results in SES are generally lower than Nabi's.

As we all know, the EC rates by SES is quite relevant for
simulations in the process of collapse and explosion in the massive
star. On the other hand, the neutrino energy loss due to EC also
plays an important role in the process of the poignant supernova
explosions. In order to understand the supernova explosion mechanism
and evolution, in order to clear the effect from cooling system of
stars by SES, more and more astronomers and physicists are
interesting in theses problem and try their best to seek the key.
 How would the SES effect on the neutrino energy loss in stars? How would the SES effect
others weak interactions in the prosess of stellar evolution? How
would the SES effect on the cooling system in massive stars? These
challenging problems will be our next objectives.

\section*{Acknowledgments}

This work was supported by the Advanced Academy Special Foundation
of Sanya under Grant No 2011YD14.

\bsp

\label{lastpage}

\end{document}